\documentstyle{l-aa}
\newcommand{\figura}[4]{
	\begin{figure}
	\picplace{#1 cm}
	\caption{#3}
	\label{#4}
	\end{figure}
}
\begin{document}
\thesaurus{ 11.05.2; 11.07.1; 11.09.4; 11.11.1; 11.19.2; 11.19.6}
\title{M=1 and 2 Gravitational Instabilities in Gaseous Disks: I. Diffuse Gas}
\author{S. Junqueira \inst{1,2} and F. Combes \inst{1}}
\institute{ DEMIRM, Observatoire de Paris, 61 Av. de l'Observatoire, 
F--75014,  Paris, France
\and CNPq-Observat\'orio Nacional, DAGE, R. Gal. Jos\'e Cristino 77, 
CEP 20921--400, Rio de Janeiro, Brazil}
\maketitle
\begin{abstract}

We report the results of self-gravitating  simulations  of 
spiral galaxies, modeled by stellar and gaseous components, developed  
to investigate in particular the role of dissipation in
the evolution of galaxy disks. The gas disk is simulated 
by the Beam-Scheme method, where it is considered as a self-gravitating
fluid. The results suggest that the gravitational coupling between 
the stars and gas plays a fundamental role in the formation and
dissolution of stellar bars, depending on the gaseous mass concentration 
and on the degree of dissipation.
In addition we remark that initially concentrated gas 
disks can be unstable to the one-armed (m=1) spiral perturbations, 
which may explain the lopsided features observed in the gas 
distribution of the late-type isolated galaxies. The
development of the m=1 feature slows down the radial gas flows towards
the center, since the large-scale gravity torques are then 
much weaker.


\end{abstract}

\section{Introduction}

  While the gravitational instabilities of a purely 
stellar disk have been widely studied, mainly through
simulations, and begin to be well understood, the situation
is much more open for a self-gravitating disk composed
of stars and gas. Even when a stellar disk is stabilised
against axisymmetric instabilities (Toomre criterion, 1964),
it can be the site of spiral and bar instability
(e.g. Sellwood \& Wilkinson 1993), together
with z-instabilities (e.g. Combes et al 1990).
These heat considerably the stellar disk, that 
can then no longer sustain spiral structure. 
Instabilities can be suppressed by reducing the
effective self-gravity of the stellar disk, either
through the addition of a hot bulge or halo (Ostriker \& Peebles 1973), 
or through disk heating, i.e. increasing the velocity
dispersion (Athanassoula \& Sellwood 1986).

The behaviour of a self-gravitating disk of
both stars and gas, even if the latter represents
only a few percents of the total mass, presents
much more variety and complexity. Due to the 
dissipation, the gaseous component remains cool,
and evacuates the heating due to the gravitational
instabilities: spiral structure can then be
continuously renewed (e.g. Sellwood \& Carlberg 1984),
even in the stellar disk. Criteria for stability
are more complex, the gravitational coupling between 
gas and stars making the ensemble unstable, even when
each component would have been separately stable
(e.g. Jog \& Solomon 1984, Bertin \& Romeo 1988, 
Elmegreen 1994). Gravity torques
exerted on the gas produce strong radial flows,
and gas can accumulate at Lindblad resonances and
form several rings (e.g. Schwarz 1981, Combes 1988).
The strong gas concentration in the nucleus can
destroy stellar bars (Friedli \& Benz 1993), after having
sometimes triggered a nuclear bar within the main one
(Friedli \& Martinet 1993). This bar destruction
can be explained by an increase of chaotic orbits in the 
stellar bar (Hasan \& Norman 1990).  The copious inflow of gas
into the nuclear region may be one of the main source of
nuclear activity in galaxies (Phinney, 1994). 

When the gas possess too much self-gravity,
however, it forms lumps through Jeans instability,
and the lumpiness of the gas can scatter
the stars, randomize their motions, and prevent
any bar formation (Shlosman \& Noguchi 1993). 
While the gas is triggering bar instability
when it represents only a few percents of the total mass,
it can play the inverse role when its mass is above $\approx$ 10\%.

All the complexity of the evolution of a self-gravitating
galaxy disk composed of stars and gas has not yet been explored;
moreover, the detailed physics of the interstellar medium
has not yet be quantified: we still do not know the 
exact amount of dissipation, or the energy balance
between dissipation and star formation, etc.. We want
to study to what extent the galaxy evolution depends
on these parameters, through the gas behaviour. We want
to compare the behaviour of the two main gas phases,
condensed and diffuse.

Observations show that the gas component is a multi-phase medium,
spanning a wide range of densities and temperatures.
To drastically simplify, we can consider two distinct forms  - 
dense, cold and clumpy molecular clouds and warm diffuse gas - with 
different spatial distribution but having comparable total masses.
The two aspects of the gas component have been studied in the
literature: ensembles of interstellar clouds that 
undergo collisions, but do not behave as a fluid (Schwarz 1981, 1984; 
Roberts \& Hausman 1984; Combes \& Gerin 1985). The collisions take place 
mainly along spiral arms. This method describes essentially  the 
molecular gas component. The response of the diffuse 
gas in barred potentials was also explored (Huntley~et~al 
1978; Roberts~et~al. 1979; Sanders \& Tubbs 1980; van Albada \& 
Sanders 1982; Prendergast 1983; van Albada 1983; Contopoulos~et~al. 
1989). In these calculations the gas was considered as a fluid,
submitted to pressure forces, and undergoing shocks. In 
most cases, the gas self-gravity was not considered. 

In the present paper we concentrate on the self-gravitating diffuse
phase, and present numerical simulations of spiral 
galaxies modeled by two components, stars and gas-fluid. The 
primary motivation is to quantify the amount of dissipation
(either artificial, turbulent or "gravitational")
and radial flow of the gas in the formation and evolution of 
the gravitational instabilities observed in spiral disks. 
In particular we discuss phenomena that can be explained by 
the star-gas coupling. Future studies will compare
the condensed with the diffuse phase behaviour, and will
study a combination of the two.
 
This paper is organized as follows. In \S~2 we discuss the 
computational methods, the galaxy model used and the initial 
conditions of our simulations. In \S~3 we present the results 
concerning the m=2 instabilities, and in \S~4 the m=1
and one-armed spiral instabilities. Section 5 quantifies the effects of
viscosity and section 6 summarizes our conclusions.

\section{Numerical Simulations}

\subsection{Computational Methods}

We simulate the self-gravity of both the stellar and gaseous components,
and solve the N-body problem by the 2D-Fast Fourier 
Transforms method. At each integration step, density (of stars
and gas) is computed at each point of a grid of $256\times256$ 
square cells. Fourier transform of density and inverse transform
of potential are then carried out. The potential is softened within
250 pc, which is also the cell size. Each particle contributes 
to the density of the four nearest grid points, according to 
cloud-in-cell (CIC) interpolation procedure (Birdsall \& Fuss 
1969). There are 36000 particles representing the stellar component. 
The gaseous component is represented by a fluid. 

The fluid is simulated according to the Beam-Scheme method 
developed by Sanders \& Prendergast (1974). This technique for 
solving the hydrodynamical equations, is an explicit Eulerian 
scheme. The gas is evolved on a $256\times256$ 2D-Cartesian grid, 
with mass, momentum, and energy densities 
being specified in each cell. The presumed Maxwellian velocity 
distribution in each gas cell is approximated by five delta 
functions (or beams), which reproduce the appropriate moments 
of the velocity distribution function. These beams are allowed to 
move over a time step transporting mass, momentum and energy into 
adjacent gas cells. Transport is taken into account to determine 
the new mass, momentum, and energy in each cell, and these moments 
are used to describe the new Gaussian velocity distributions 
for each cell of the gas. 

\subsection{Galaxy Model}

As for the initial conditions, the model galaxy is composed of a
stellar disk, an analytical spherical bulge and a gaseous disk. 
The stellar particles are initially distributed according to a
Toomre disk (Toomre, 1962) of mass $M_d$ and radial 
scale length $a_d$. The gaseous component of mass $M_g$ is modeled by another 
Toomre disk, of radial scale length $a_g$, with velocities 
distributed in beams according to the Beam-Scheme method. A 
Plummer potential (Binney \& Tremaine 1987) of radial scale 
length $a_b$, represents the bulge. In all simulations the bulge 
and the disk have the same mass ($M_d=M_b$) and the gas mass corresponds 
to 10\% of the total mass of the galaxy ($M_g=0.1M_t$; 
$M_t=10^{11}~M_{\sun}$). The initial cutoff radius of the stellar 
and gaseous disks are 12~kpc and 16 kpc, respectively. The 
characteristic scale of the stellar disk, of the bulge and of the 
gaseous disk are free parameters. Their values are shown in 
Table~1. Initially, the stellar particles have a velocity dispersion 
as a function of radius corresponding to the Toomre parameter $Q=1$
all over the disk. The rotational velocity is corrected for
asymmetric drift.
 The units of time, distance and mass for all 
simulations are $10^{7}$~years, 1~kpc and $4\times10^9~M_{\sun}$ 
respectively. All the experiments have been evolved through a total time 
equal to $2.2\times10^{9}$~years. The time-step is 0.1 unit= 1 Myrs.
  
In the experiments A, B, C and D the gaseous component is not 
included. These are control simulations in order to probe the
gas influence. In the E, F, G and 
H experiments the stellar components are described by the same 
initial conditions as in the A, B, C and D experiments but a gaseous 
disk, of scale length $a_g=6$~kpc, is included. A comparative 
analysis of these experiments allow us to draw important 
conclusions about star-gas coupling. Experiments I and J test the
role of the initial concentration of the gaseous disk in the 
evolution of the system.

\begin{table}
\begin{flushleft}
\caption{Parameters of the Experiments}
\begin{tabular}{c c c c c}
\hline
Exp. & $a_d$ & $a_b$ & $a_{g}$\\
\hline
A & 3.0 & 1.0 &     \\
B & 4.5 & 1.0 &     \\
C & 3.0 & 2.0 &     \\
D & 4.5 & 2.0 &     \\
E & 3.0 & 1.0 & 6.0 \\
F & 4.5 & 1.0 & 6.0 \\
G & 3.0 & 2.0 & 6.0 \\
H & 4.5 & 2.0 & 6.0 \\
I & 3.0 & 1.0 & 4.0 \\
J & 3.0 & 1.0 & 8.0 \\
\hline
\end{tabular}
\end{flushleft}
\end{table}

\section{M=2 instabilities}

In this section we discuss the role of gas dissipation
in the formation and maintenance of gravitational 
instabilities ($m=2$ spirals, bars).

\subsection{Influence of the gas }

In figure \ref{f1} we present the linear isodensity contours 
of the stars for the experiments A, B, C and D at the 
time t=140 (10$^7$ yrs). In these four 
experiences, as described in the preceding section, the gaseous 
component is not included.  We can note that in spite of the large bulge
mass ($M_d=M_b$) a central bar develops. The spiral arms 
formed initially are only transient. They help to transfer angular
momentum outwards to make the bar grow.
The characteristic time-scale for bar formation in these 
experiments varies according to the degree of concentration 
of the initial bulge and stellar disk. Models with a smaller
bulge scale length $a_b$, developped their bar in a shorter time-scale
than those with a larger initial value of $a_b$. This can be
checked by comparing experiments A and C or 
the experiments B and D, and can be interpreted by a shorter
dynamical time-scale in the center of A and C. The bar is also shorter 
in length in Exp. A and C, and the corotation radius is smaller.
 Considering now the same 
initial mass concentration for the bulge (experiments A and B) 
we verify that disks initially more concentrated (experiment A) 
allow a more rapid bar formation. This could be due to the
larger self-gravity of the stellar disk in the center, together with the
shorter $t_d$ there. 

The main conclusion from these four experiments is
the high longevity of the bar: the bar kept its amplitude 
up to the end of simulations (2.2 Gyrs). 
In the absence of gas, bars are long-lived,
even in the presence of a concentrated bulge.

On the contrary, a great variety of scenarios can occur when
the gaseous component is included.  In figures \ref{f2}, 
\ref{f3}, \ref{f4} and \ref{f5} we present the linear 
isodensity contours of the stars and gas for the 
experiments E, F, G and H, at four different epochs (t=20, 
80, 140 and 200 $\times$ 10$^7$ yrs). In these experiments,
the stellar and gaseous components can develop a 
bar (experiment E), in the same way as pure stellar runs,
or evolve to various degrees of central mass 
concentration (Exp. F, G and H). 
 In that case either the bar forms, and is then
destroyed rapidly (experiments G and H) or even the bar
does not form at all (experiment F).
 
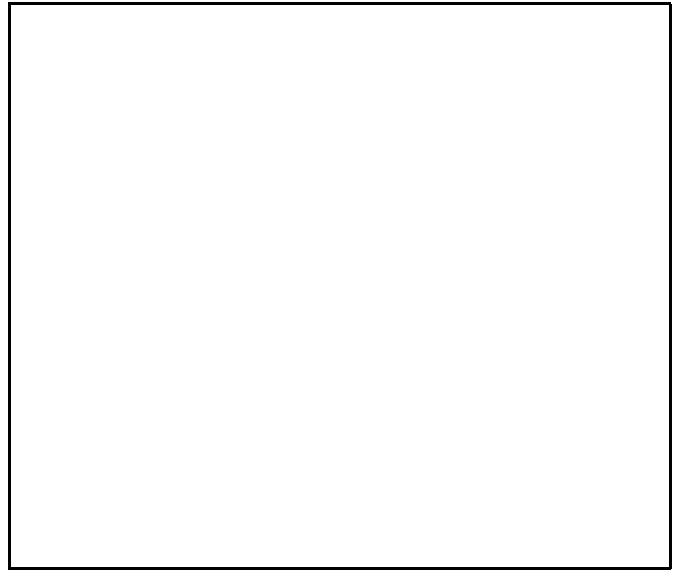
\figura{7.5}{fig1}{Exp. A, B, C and D: linear isodensity 
contours of the stars at epoch t=140 (in 10$^7$ yrs). 
X and Y are in kpc.}
{f1}

\figura{7.5}{fig2}{Exp. E: linear isodensity contours 
of the stars (left side) and of the gas (right 
side) at epochs t=20, 80, 140 and 200 (in 10$^7$ yrs). 
X and Y are in kpc.}
{f2}

\figura{7.5}{fig3}{Exp. F: linear isodensity contours 
of the stars (left side) and of the gas (right 
side) at epochs t=20, 80, 140 and 200 (in 10$^7$ yrs). 
X and Y are in kpc.}
{f3}

\figura{7.5}{fig4}{Exp. G: linear isodensity contours 
of the stars (left side) and of the gas (right 
side) at epochs  t=20, 80, 140 and 200 (in 10$^7$ yrs). 
X and Y are in kpc.}
{f4}

\figura{7.5}{fig5}{Exp. H: linear isodensity contours 
of the stars (left side) and of the gas (right 
side) at epochs  t=20, 80, 140 and 200 (in 10$^7$ yrs). 
X and Y are in kpc.}
{f5}

\subsection{Bar amplitude and pattern speed}
  
As described above, the bar strength and evolution depends
on the central mass concentration and the gas fraction.
We focus in this section on the bar evolution,
the results of the experiment F where a central mass 
concentration is obtained without the formation, even transient, 
of a bar will be discussed in the next section.

To better identify the $m=2$ perturbation, we have computed, 
for all simulations, the angular velocity of the pattern 
$\Omega_p$ as a function of radius. A mode would be 
characterized by a constant $\Omega_p$ over a range of radii. 
A Fourier analysis  of the density distribution or of the 
potential allows to decompose the perturbation present in the 
galaxy in several components (m). The index m indicates the 
symmetry of each component or, in the case of spirals, the 
number of arms. The analysis provides, at each epoch, both the 
strength or amplitude of the m component, which growth 
rate can then be estimated and its phase, from which $\Omega_p(r)$ 
can be derived. We have computed the growth rate for the 
spiral-bars in the experiments A, E, G and H assuming exponentially 
growing modes. The behavior of the bar amplitude, the maximum ratio of the
$m=2$ tangential force to the radial force ($p_2$) 
as a function of time is shown in figure \ref{f6}. 
While for the experiments A and E $p_2$ remains approximately constant 
as in a steady-state (although the bar weakens somewhat
in experiment E), in  
experiments G and H it decreases rapidly. The growth time-scales
obtained for the bars are approximately 7~$t_d$ (Exp. A), 6~$t_d$ 
(Exp. E), 5~$t_d$ (Exp. G) and 4~$t_d$ (Exp. H) where $t_d$ is 
the dynamical time scale ($\approx 4\times10^7$ yrs). We remark 
that in Exp. A and E the behavior of $p_2$ in the first 
instants of the simulation (before the exponential bar growth) 
indicates the presence of transient spiral structure
preceding bar formation. We have calculated for the 
experiments A and E the angular velocity (or pattern speed) 
$\Omega_p(r)$ for the bar perturbation. In figure~\ref{f7} 
are shown the $\Omega$ (solid curve), $\Omega + \kappa/2$ and 
$\Omega - \kappa/2$ (dashed curves) frequency curves, where 
$\kappa$ represents the epicyclic frequency and $\Omega$ is the 
angular velocity of the galaxy. The expected 
position of the Lindblad resonances are indicated at 
the top. It can then be noted that the
bars are confined within their corotation radius in both 
experiments. 
 
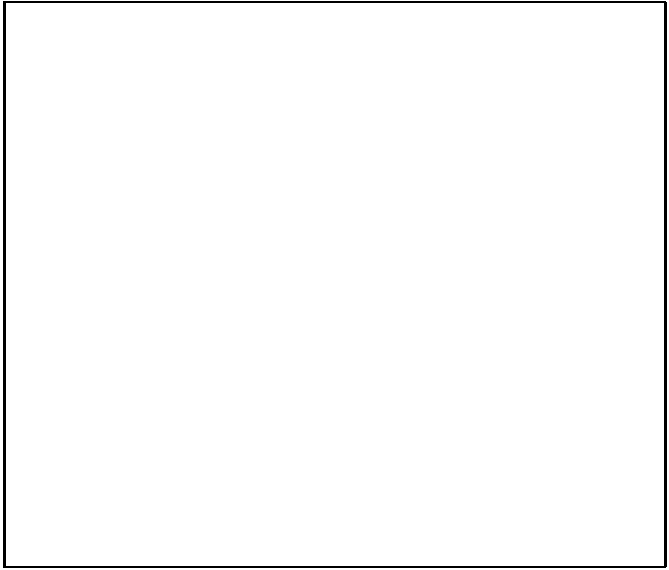
\figura{7.5}{fig6}{The strength of the bar perturbation
$p_2$ as a function of time (in 10$^7$ yrs units) for the 
experiments A, E, G and H. The m=2 amplitude has been averaged over 
radius 0 to 5kpc.}
{f6}

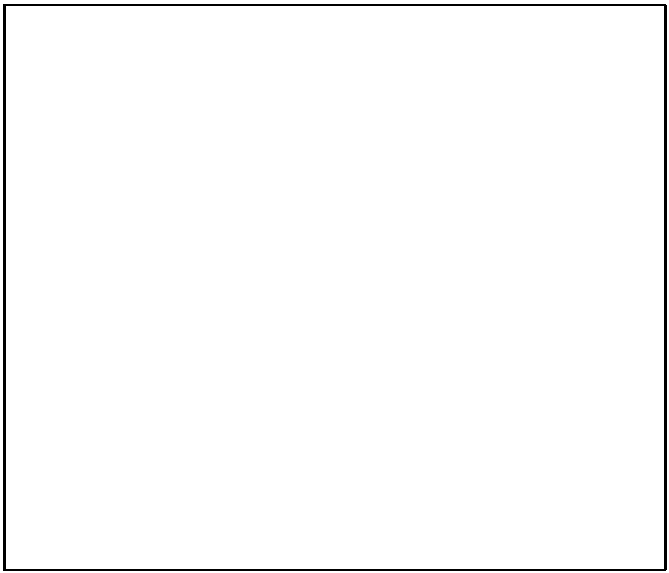
\figura{7.5}{fig7}{The frequency curves (in units of 
100~km/s/kpc) as a function of radius (in kpc) and the 
position of the pattern speed $\Omega_p(r)$ for 
experiments A (top) and E (bottom).}
{f7}

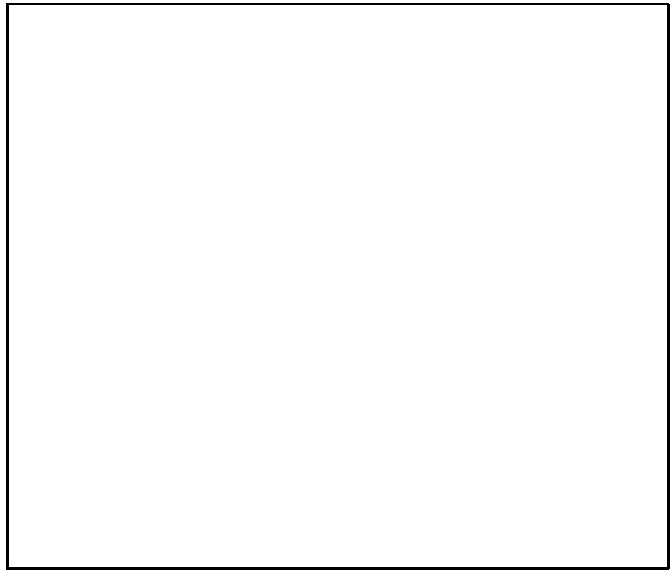
\figura{7.5}{fig8}{Exp. E, F, G and H : surface density 
(in code units) of the stars and gas as a function of 
radius (in kpc) at four different epochs t=20, 80, 140 and 
200 (in 10$^7$ yrs). The letters at the bottom left corner of each 
figure indicate (s) stars and (g) gas.} 
{f8}

Figure~\ref{f8} displays the evolution of the surface density of 
stars (left side) and gas (right side), as 
a function of radius, at four different epochs.
 These quantify the consequences of radial gas flows 
and mass concentration of stars and gas in 
the central region. Radial flows correspond to angular momentum
transfer driven by non-axisymmetric gravitational instabilities,
and the subsequent central mass concentrations react back on 
gravitational instabilities.

As is apparent in figure~\ref{f8}, in all simulations
the surface density of stars still follows, 
at the first epoch (t=20), the imposed initial conditions.
On the contrary, the gas surface density is determined at
the first epoch, by the bulge mass 
concentration. Such behavior determines the 
future evolution of the stellar density distribution. In other 
words, at t=20 the gas, in the experiments E and F ($a_b=1$), 
is more concentrated than in the experiments G and H. This strong 
initial concentration in experiment E is the cause of a high central
accumulation of stars, and after some time favors the formation and 
maintenance of the bar. More precisely, it avoids a violently
unstable solution, that will heat the disk and destroy the bar.
In experiments G and H the bar forms more rapidly 
but it is not maintained. At t=80, 140 and 200 we observe 
that the gas in experiments E, G and H can reach the same 
maximum of density. While in experiment E the region out of 
the bar is depopulated from its gas, the contrary occurs
in experiment F. In this experiment a strong central 
concentration of gas at the initial epochs together with a 
considerably lower concentration of stars induce the 
formation of an m=1 spiral mode in the central region of the galaxy.
The concentration of gas and stars is then almost completely stopped.

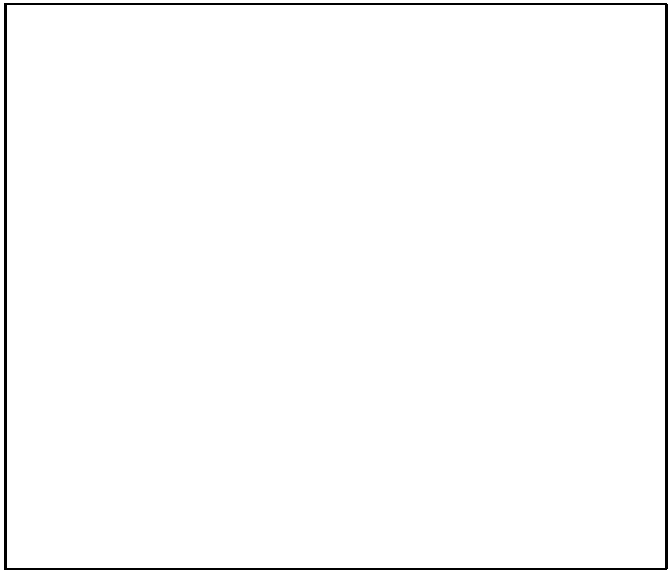
\figura{7.5}{fig9}{Exp. E : the phase (in radian) of m=2 
Fourier component for the stellar (left side) and the 
gaseous (right side) surface density and the total potential 
as a function of radius (in kpc) at four different epochs.
Also shown is $d_2$, the amplitude of the m=2 component of the surface
density of stars and gas. The scale of $d_2$ is in the right side.}
{f9}

We can observe that although the bar obtained in Exp. A 
shows approximately the same growth rate as that obtained 
in Exp. E, it is relatively longer. This is certainly due
to the larger mass concentration of Exp. E, due to the gas
dissipation. Also spiral structure is present in Exp. E in contrast
to A, in the gas as well as in the stars.
We remark also that there is a phase shift 
between the gas bar (leading) and that of the stars of 
approximately 30\degr in the first epochs. This explain
the strong gravity torques exerted by the potential on the gas,
which is driven inwards (cf Combes 1988). In her 
study about the secular evolution of spiral galaxies, Zhang (1995), suggested 
the existence of a collective dissipation mechanism responsible 
for the secular evolution of the stellar disks of spiral galaxies. 
Since the disk is not in complete equilibrium, there exists also a
phase shift between the self-consistent spiral potential 
and the stellar density. This phase shift indicates that there is 
a torque applied by the spiral potential on the stars,
and a secular transfer of energy and angular momentum between 
the stellar disk and the spiral density wave. In figure~\ref{f9} 
we show, for experiment E, the phase of the m=2 Fourier 
component for the gas and stellar surface densities and for the total potential 
as a function of radius at four different epochs, t=20, 80, 140 and 200 
(in 10$^7$ yrs). At the beginning (t=20) gaseous and 
stellar surface densities are in phase with the potential. 
See also in figure~\ref{f2} the agreement between the spiral 
arms developed in the gas and in the stars. At t=80, 140 and 200 there 
is in the gas a phase-shift between the surface density and 
potential, implying a large torque felt by the gas
for a large range of the radii; after corotation, the stellar 
surface density leads the potential, while just before CR
the potential leads the two components. Corotation is where
the potential phase shows the maximum gradient.
At t=200, a gas spiral feature just outside corotation introduces
important phase shifts, but they are not related to the bar,
as shown by $d_2$ the m=2 amplitude in the surface density.
 
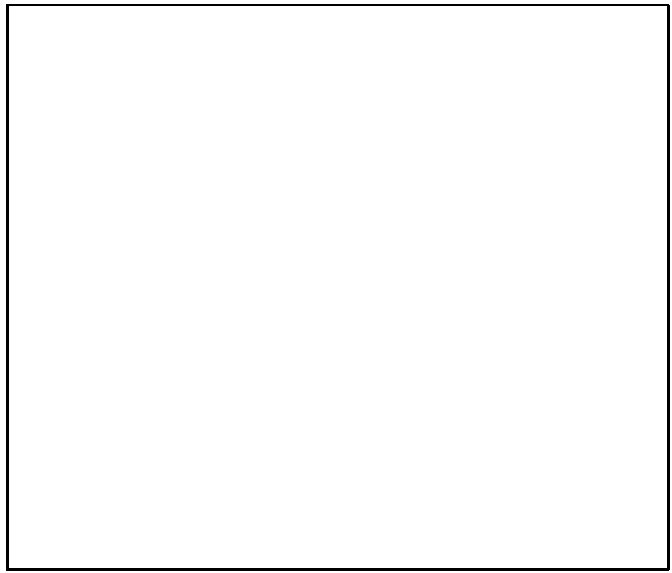
\figura{7.5}{fig10}{Exp. I, E, and J: linear isodensity 
contours of stars and gas at epoch
t=140 (in 10$^7$ yrs). X and Y are in kpc.}
{f10}

\figura{7.5}{fig11}{Exp. E, I, and J : surface density 
(in code units) of stars and gas as a function of 
radius (kpc) at four different epochs, t=20, 80, 140 and 
200 (in 10$^7$ yrs). The letters at the bottom left corner of each 
figure indicate (s) stars and (g) gas.} 
{f11}

In order to probe the conditions required for the formation of 
a bar (as in experiment E) or an m=1 spiral mode 
(Exp. F), two other experiments (I and J) were done. In 
these experiments we used the same initial conditions 
for the stellar disk as in experiment E 
but the gaseous component has different degrees of 
concentration determined by the characteristic 
radius $a_g$. In figure~\ref{f10} we compare the linear 
isodensity contours of the stars and gas at the 
time t=140 for Exp. I, E and J: when
the gaseous disk is initially less concentrated a bar can  
form (Exp. J and E) while when it is initially more concentrated 
(Exp. I, $a_g=4$) an m=1 spiral mode develops. In this latter 
case the stars and gas concentrations show a final behavior very 
different from that observed in the bar experiments. The 
results can be interpreted comparing the 
surface density of stars and gas 
at different epochs as shown in figure~\ref{f11}. 
Although the stellar density of Exp. I does not stand out from other 
experiments at any time, the gas density is different.
While in Exp. E and J the 
bar forms, the concentration of gas increases and the region out 
of the bar is depopulated, in Exp. I, the development of an m=1 spiral mode 
slows down the gas concentration. We discuss now in more details the
formation of these m=1 spiral modes.

\section{M=1 instabilities}

Eccentric asymmetries in the distribution of light in spiral galaxies 
have been known for a long time (Baldwin et al. 1980; Wilson 
\& Baldwin 1985). Lopsided features are preferentially observed 
in the distribution of gas in late-type spiral galaxies. 
In several cases these features can be identified as one-armed 
spirals (m=1 mode). More frequently, nuclei of galaxies are observed
displaced with respect to the gravity center, as in M33 and M101
(de Vaucouleurs \& Freeman 1970). Barred spiral galaxies can have
their kinematical center displaced from the bar center (Christiansen \&
Jefferys 1976, Marcellin \& Athanassoula 1982, Duval \& Monnet 1985). The
nucleus of M31 reveals such an off-centring (Bacon et al 1994) which 
has been interpreted in terms of an m=1 perturbation (Tremaine 1995).
Miller \& Smith (1992) have studied through N-body simulations of
disk galaxies, a peculiar oscillatory motion of the nucleus with respect
to the rest of the axisymmetric galaxy. They interpret the phenomenon
as an m=1 instability, a density wave in orbital motion around the center
of mass of the galaxy. Weinberg (1994) shows that a stellar system can
sustain weakly damped m=1 mode for hundreds of crossing times. A fly-by
encounter could excite such a mode, and explain off-centring in most spiral
galaxies, or the forcing of a halo could maintain a time-dependent potential
in the galaxy (Louis \& Gerhard 1988).
 
In this section we concentrate on perturbations
with azimuthal wavenumber m=1. The results of one of these 
simulations (Table~1, experiment F), which shows gravitational 
instabilities of this nature, is presented in figure~\ref{f3}. 
The linear isodensity contours of the gas show features in the 
central region that can be identified as an m=1 spiral mode.
The Fourier analysis allows us to reconstruct the density distribution of the
gas from the combination of the various components and single out the 
contribution of each one in the final configuration.
In figure~\ref{f24} we present the linear isodensity contours of the
gas distribution reconstructed from the two main components
(m=1 and m=2). Comparing figures \ref{f3} and \ref{f24} we can 
check the relative contribution of these components and how they 
add to each other. 

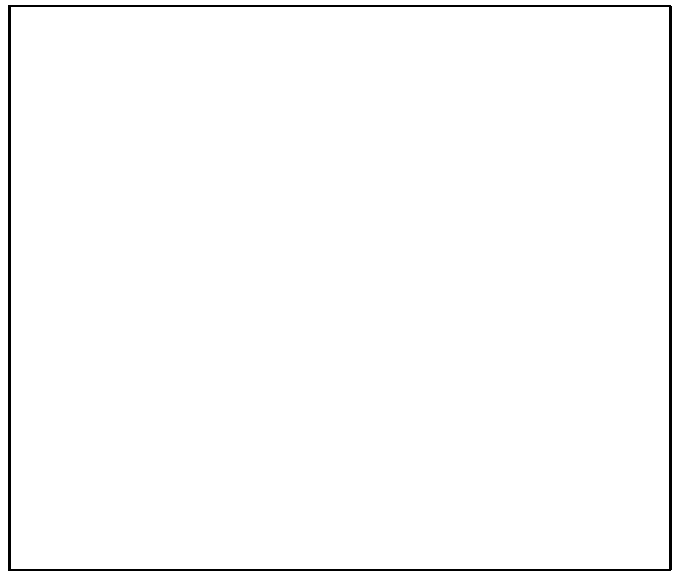
\figura{7.5}{fig24}{Exp. F: linear isodensity contours do gas reconstructed 
from the m=1 and m=2 Fourier components epochs t=20, 80, 140 
and 200 (in 10$^7$ yrs). X and Y are in kpc.}
{f24}

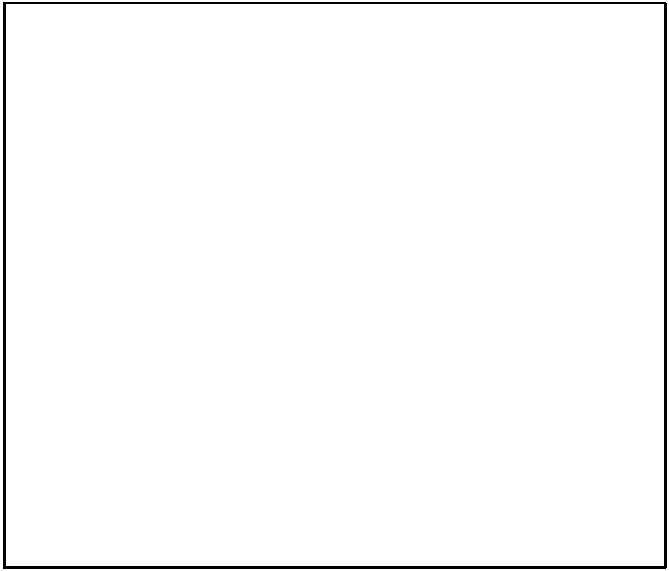
\figura{7.5}{fig12}{Exp. F: The angular velocity $\Omega_p(r)$ of the 
m=1 wave and the frequency curves (in units of 100~km/s/kpc) as a 
function of radius (in kpc).}
{f12}

For this experiment, we have calculated the 
angular velocity $\Omega_p(r)$ of the m=1 wave.
The m=1 wave is confined in the central region 
(r $\leq$ 3 kpc), with a pattern speed of the
order of 400km/s/kpc. Figure~\ref{f12} displays the 
$\Omega$ (solid curve), $\Omega + \kappa$ and $\Omega - \kappa$ 
(dashed curves) frequency curves, together with the
expected position of the resonances (CR and OLR).
The m=1 perturbation develops between the center 
of the galaxy and its OLR radius, slightly above 3 kpc. 
 Outside this radius, independent perturbations, of symmetries
m=1 and 2, develop (see figure~\ref{f21}. The pattern speed of these 
is of the order of 10-20 km/s/kpc. In figure~\ref{f21} we show the
linear isodensity contours of the gas in Exp. F at twelve consecutive epochs
(from t=140 to 142.75, in 10$^7$ yrs). We can note that the central region 
(about 3 kpc around the densest cell) 
of the m=1 spiral wave rotates rapidly, completing almost two rotations
during this time interval while the outer spiral structure evolves more slowly.
 The value of the central $\Omega_p(r)$, considerably greater than 
in the outer regions of the galaxy, precludes any non-linear coupling 
of the patterns in the two regions.
In figure \ref{f23} we show the same linear isodensity contours displayed 
in figure \ref{f21} rotated with a constant angular velocity 
($\Omega_p(r)$=400km/s/kpc). The position of the m=1 wave at each time in the  
central region confirms that the pattern speed is in fact considerably 
greater there. 

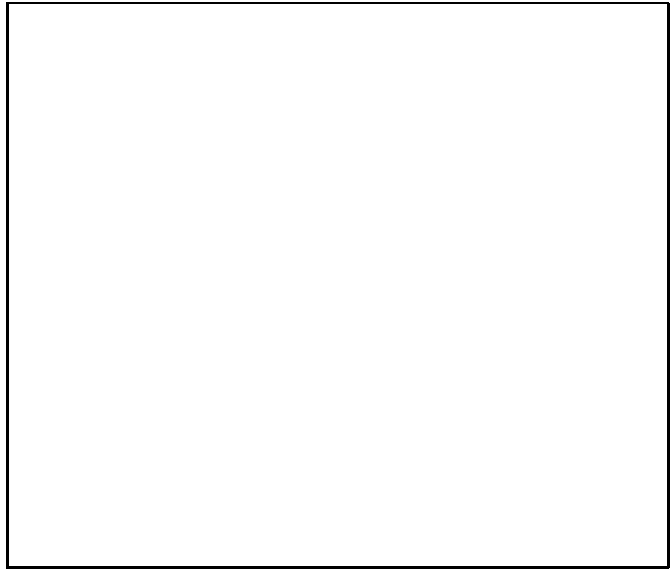
\figura{7.5}{fig21}{Exp. F: linear isodensity contours of the gas at 12
close-by epochs (in 10$^7$ yrs). The circles show the region of 3 kpc around 
densest cell. The axis X and Y are in kpc.}
{f21}

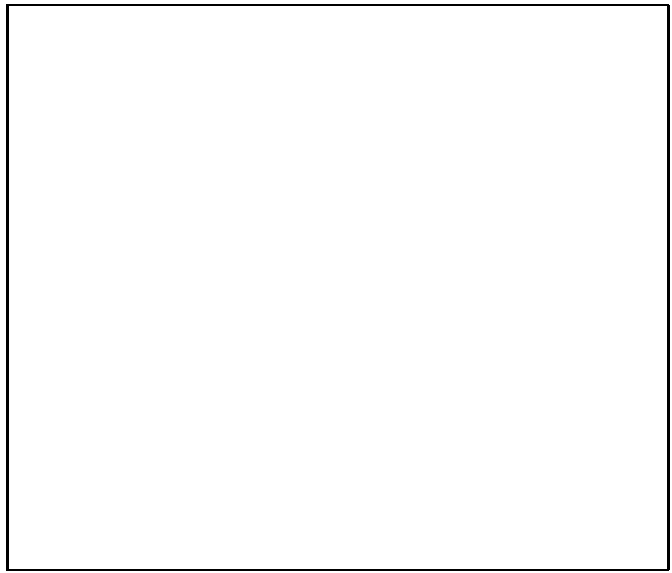
\figura{7.5}{fig23}{The same linear isodensity contours of the gas at 12
close-by epochs (in 10$^7$ yrs) shown in figure 14 rotated of 
$\Omega_p(r)=$ 400km/s/kpc.}
{f23}

It is important to remark that, at the initial instants of 
the simulation (t=20) in the central region (r=3~kpc), the mass 
of the gas increases to approximately 6 times its
initial value, just before the appearance of the m=1 mode.
After that, the growth rate of gas mass in this region slows down
steadily with time. In the other experiments where the m=1 does not occur,
the gas mass continues to grow rapidly.
The slowing down of radial gas flow coincides also with 
a stabilisation of the gas angular momentum in the center,
as soon as the m=1 arm is identified.
 
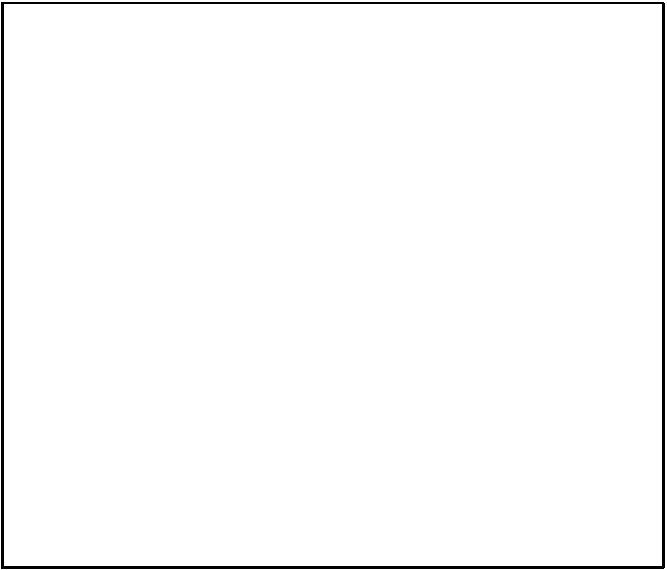
\figura{7.5}{fig13}{Exp. F: position of the center of mass of the 
gas (circles) and the center of mass of the stars (*) relative 
to the center of mass of the system (crosses) calculated within
radii from r=1 kpc to 10 kpc at different epochs indicated at the top left 
of each figure (in 10$^7$ yrs). X and Y are in kpc.
The position of the center of mass of the gas and of the stars 
calculated within the same radius are connected by a straight line.}  
{f13}

\figura{7.5}{fig14}{Exp. E: position of the center of mass of the 
gas (circles) and the center mass of the stars (*) relative 
to the center of mass of the system (crosses) calculated inside 
radii from r=1 kpc to 10 kpc at different epochs indicated at the top left 
of each frame (in 10$^7$ yrs). X and Y are in kpc.}  
{f14}

Adams et al (1989) have discovered numerically 
similar gravitational instabilities in 
gaseous disks associated with young stellar objects: eccentric
m=1 modes, when the star did not lie at the center
of mass of the system. Shu~et~al. (1990) presented 
an analytical description of a modal mechanism, the SLING amplification,
or "Stimulation by the Long-range Interaction of Newtonian Gravity". 
This mechanism uses the corotation amplifier, where the birth of positive 
energy waves outside strengthens negative energy waves inside. In the SLING
mechanism, a feedback cycle is provided by four waves outside
corotation; long-trailing
waves propagate from the OLR inward to the Q-barrier at CR where they refract
in short trailing waves. These propagate outwards, cross the OLR, and
reflect back at the outer edge of the disk. This is a critical point,
the whole amplification mechanism depends on the reflecting character
of this outer edge. Then a short leading wave propagates inwards from
the edge, through OLR, towards CR, where it refracts again into a long
leading wave, that is then reflected at OLR. An essential point here
is also the ability of waves not to be absorbed at OLR, that is why this
mechanism apply to gaseous disks only.  Using 
a WKBJ analysis, they derived from the dispersion relation, and
the condition of a constructive reflection, the required
pattern speed for these modes. The instability arises uniquely from the
displacement of the central star from the center of mass of the 
system, which creates
an effective forcing potential. In our simulation the same effect 
is occurring, i.e. the center of mass of the gas and of the 
stars is displaced from the center of mass of the system and 
they are displace in opposite positions, inducing the m=1 wave 
formation. In other words, the modal mechanism proposed by 
Shu~et~al. (1990) may be in action in our experiment. In 
figures \ref{f13} and \ref{f14} we show the position of the center 
of mass of stars and gas calculated within radii 
from r=1 kpc to r=10 kpc, at 12 different epochs (indicated at 
the top left of each frame) for experiments F and E
respectively. In this figures the circles indicate the position 
of the center mass of the gas, the symbol (*) the position 
of the center mass of the stars and the crosses the center 
mass of the system. From the comparison of figures \ref{f13} 
and \ref{f14}, we note that when there is a bar (Exp. E) the position of 
the center of mass of the gas  and of the stars agree with the 
position of the center mass of the system, while when an m=1 spiral 
wave is identified, the center of mass of the gas and of the stars 
are in opposite locations relative to the center of mass of the system.

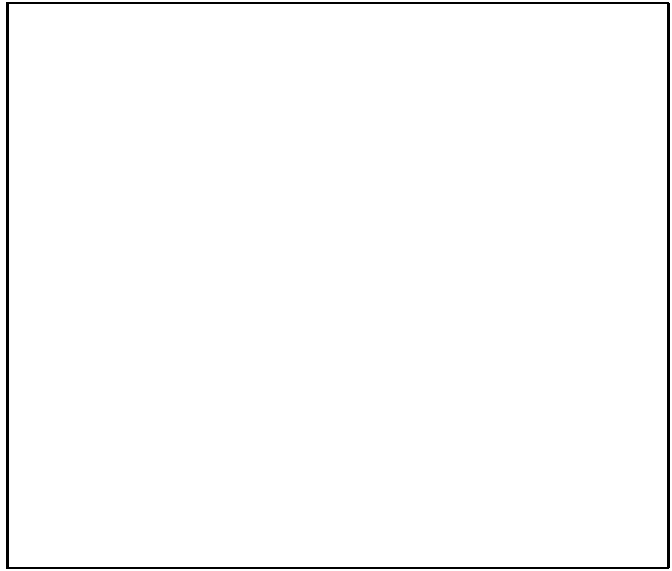
\figura{7.5}{fig22}{Exp. F: Gray-colour maps do pattern speed $\Omega_p(r)$ 
(km/s/kpc) of the m=1 spiral wave determined from the total potential 
and from the surface density of the gas as a function of the radius (in kpc).} 
{f22}

The power spectrum analysis, using the amplitude and the phase of the Fourier
components
of the total potential and of the surface density of the gas and of the stars,
is another method that allow us to evidence the presence of the spiral modes 
in our simulations.  This analysis give us the pattern speed of all present
spiral perturbations and their relative positions. 
The figure \ref{f22} displays gray-colour 
maps of the obtained angular velocity, from the potential and from the surface 
density of the gas (Exp. F) as a function of the radius, for the m=1 mode,
using a such method.  In these figure we can observe that the 
m=1 mode ($\Omega_p(r)$=400km/s/kpc), identified in the distribution of the 
gas density, it is more pronounced between the centre and the radius 3 kpc, 
extending however until the edge. The same behaviour can be verified 
in this figure for the potential, indicating that although this mode 
be more easily identified
in the central region of the galaxy, it produces a perturbation in the total
potential that extends until the edge. 

\figura{7.5}{fig15}{Exp. F: the strength of the m=1 spiral 
perturbation $p_1$ as a function of time (in 10$^7$ yrs).}
{f15}

Figure \ref{f15} presents the strength of the m=1 
($p_1$) mode as a function of time. From this,
we have computed the growth rate of the m=1 wave in the 
central region, assuming an exponentially growing mode. The 
growth time-scale obtained is approximately 3 $t_d$, at the radius 
of 3 kpc. This corresponds to the predictions of the
SLING mechanism (Adams~et~al. 1989, Shu et al. 1990).
Also, the fact that the m=1 behaviour described here occurs only when the
gas mass fraction is high, is compatible with this mechanism,
which is essentially gaseous in its origin.

\section{Viscosity}

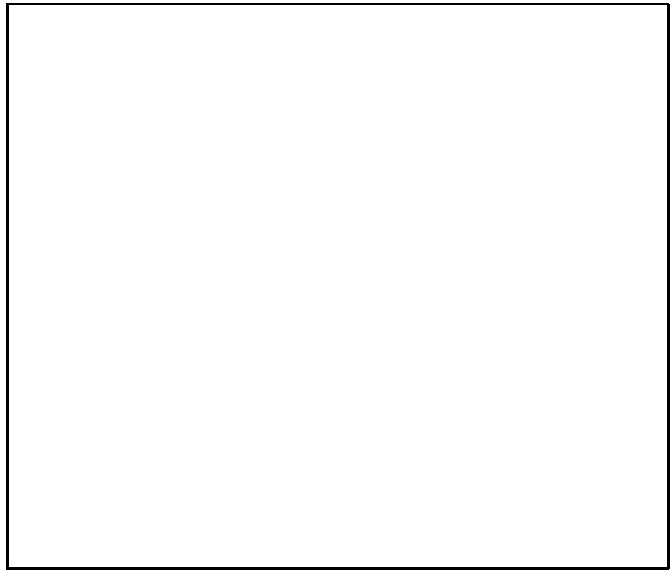
\figura{7.5}{fig16}{Exp. F1: linear isodensity contours of the 
stars (left side) and gas (right side) at epochs
t=20, 80, 140 and 200 (in 10$^7$ yrs). X and Y are in kpc.}
{f16}

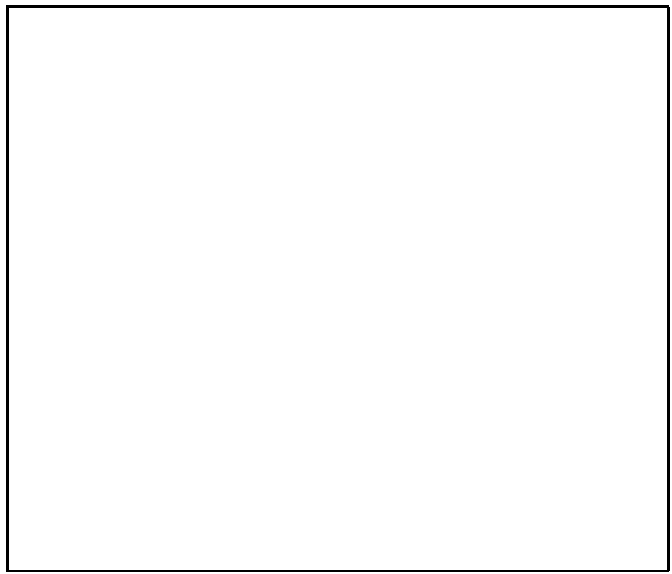
\figura{7.5}{fig17}{Exp. F1: the strength of the m=1 spiral 
perturbation $p_1$ as a function of time (in 10$^7$ yrs).
The m=1 strength has been averaged over radii 0 to 3 kpc. }
{f17}

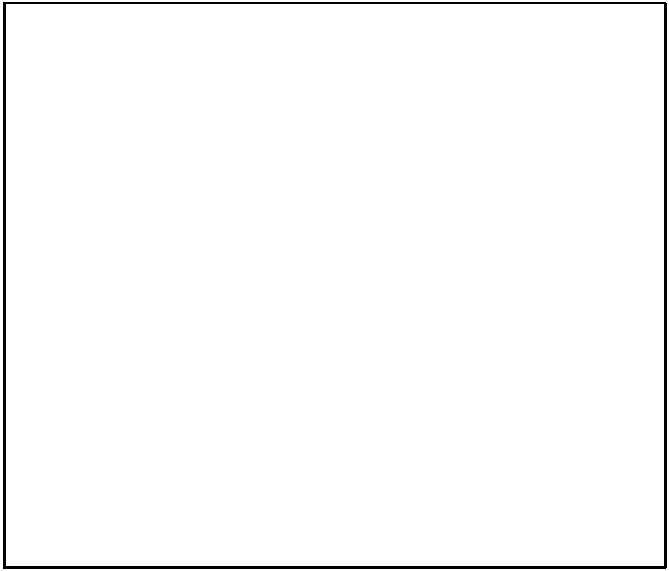
\figura{7.5}{fig18}{Exp. F and F1 : surface density 
(in code units) of stars and gas as a function of 
radius (kpc) at four different epochs t=20, 80, 140 and 
200 (in 10$^7$ yrs). The letters at the bottom left corner of each 
frame indicate (s) stars and (g) gas.} 
{f18}

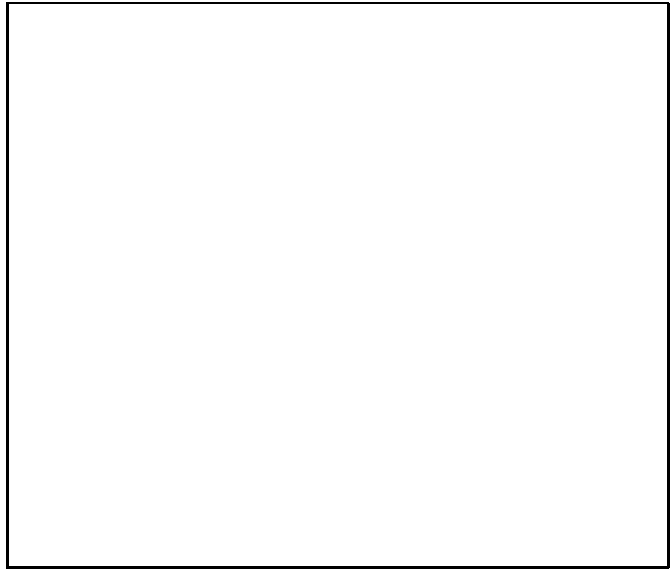
\figura{7.5}{fig19}{Exp. F and F1 : radial velocity (top) and 
equivalent viscosity of the gas as a function of time (bottom), 
at radii 3, 7 and 11~kpc.} 
{f19}

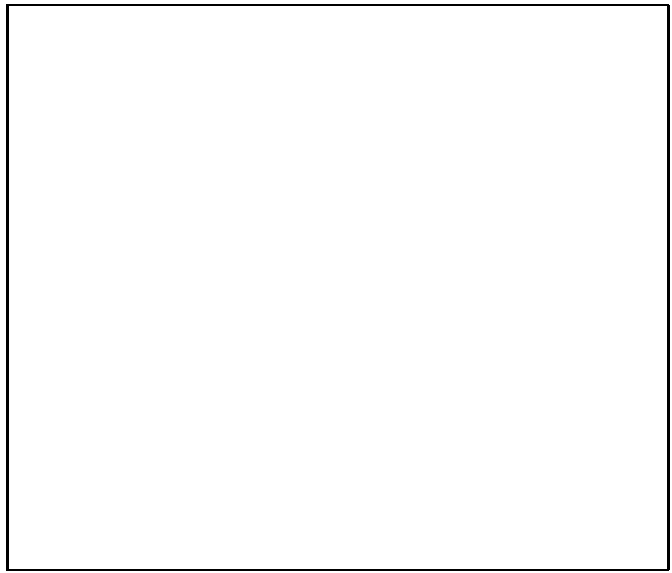
\figura{7.5}{fig20}{Exp. F and F1 : radial velocity (top) and 
equivalent viscosity of the gas as a function of radius (bottom), 
at epochs t=20, 80, 140 and 200 (in 10$^7$ yrs).} 
{f20}

The dynamical role of the dissipative gas component is essential
to account for the variety of structures observed in spiral galaxies.
 But the actual amount of dissipation is not well known, and
 worries have been expressed that the numerical treatment of the
gas component could introduce artificial angular momentum transfer, through
a too large viscosity. In the real interstellar medium, the 
viscosity is due to macroscopic turbulent motions, that can be supersonic
with a 10km/s velocity dispersion (e.g. Dickey et al 1990). 
But the corresponding viscous torques are quite weak at large-scale,
and the time-scale for viscous transport of the galactic gaseous disk
is larger than a Hubble time (e.g. Lin \& Pringle 1987, Combes 1991).
Gravitational torques are always dominant, and can transfer angular
momentum on time-scales much shorter than the Hubble time. Fortunately,
this makes gas-dynamical simulations more independent of the adopted
physics of the interstellar medium.

We therefore would like to check that the beam-scheme used here
does not introduce too large viscous torques.
In order to evaluate quantitatively the numerical viscosity of 
the method we have carried out further control
experiments. The artificial viscosity must be 
proportional to the size of the grid cell where the surface
density is computed.  In the experiment F1 we have used the same 
initial conditions as in experiment F but the grid is 
now composed of four times more cells ($512\times512$ cells) 
with four times smaller surface each. In figure \ref{f16} we present 
the linear isodensity contours of the stars and gas for 
Exp. F1 at four different epochs. As described above for the 
former experiments, we calculate for experiment F1 the growth 
rate of the m=1 spiral mode from the time evolution of its amplitude
$p_1$ plotted in figure \ref{f17}. The corresponding growth time-scale
is approximately 23~$t_d$, i.e. in this experiment the m=1 
spiral mode needs a much longer time-scale to develop, compared to Exp. F.
The analysis of the surface density distribution of stars 
and gas as a function of radius and at different epochs, 
as shown in figure \ref{f18}, shows that although 
initially (t=20) the viscosity is larger in Exp. F, 
driving a higher gas flow towards the center, at following 
epochs (t=80) the surface density distributions of Exp. F and F1 evolve 
in a similar manner. After that, from t=100, their behavior 
diverge from one another. At t=140, the gas density is more 
concentrated in Exp. F1, although the numerical viscosity is there
lower. This must be due to the fact that the m=1 arrived later,
so the gravity torques had more time to act than in Exp. F. 

In order to interpret this behavior, we calculated the 
equivalent "gravitational viscosity" of the gas 
as proposed by von Linden et al. (1995). From this point of view, the 
combination of gas dissipation and gravity torques from non-axisymmetric 
gravitational instabilities developing in the disks, is considered
as an equivalent "gravitational viscosity". The gravity torques produce an 
angular-momentum re-distribution and an evolution of the surface 
density of the gas. By monitoring this surface density one can derive 
the equivalent viscosity quantitatively. This "gravitational viscosity" 
does not correspond to a classical gaseous viscosity, but it is a 
measure of the efficiency with which gravity torques transfer angular 
momentum outward and matter inward in the system. In figure \ref{f19} 
is plotted (at top) the behavior of the radial velocity of the gas 
as a function of time at three different radii (3, 7 and 11~kpc) 
for experiments F and F1 and (at bottom) the behavior of the 
corresponding "gravitational viscosity". 
For experiment F the viscosity has a positive value initially, just
before the development of the m=1 perturbation, and converges later on towards 
a zero mean value. In experiment F1, the viscosity remains large for
a longer time, which corresponds to the delay in the
development of the m=1 perturbation with respect to Exp. F. 
To better visualize the viscosity 
behavior in these experiments, we show in figure \ref{f20} the 
radial velocity and the gravitational viscosity as a function of radius at four 
different epochs. While the viscosity in Exp. F is 
negative at the center and tends to zero at larger radii after 
the development of the m=1 mode, in experiment F1 the viscosity 
remains positive.

In summary, although the numerical viscosity is slightly higher in Exp. F,
this is largely compensated by the larger
gravitational viscosity observed in Exp. F1. When the grid cells are smaller,
the gas can condense in cold clumps more easily (see fig \ref{f16}), and the 
lumpiness of the gas has a significant gravitational effect.
Although the numerical viscosity in itself does not alter significantly the
results, the dynamic range and spatial scale of gaseous condensations,
and therefore the physics of the ISM, plays a significant role.

\section{Summary and Conclusions}

We have investigated the role of gas dynamics in the gravitational 
instabilities of galaxy disks, through self-gravitating simulations
of the coupled stellar and gaseous components. In the present
paper, we have focussed on the diffuse gas component, that 
can be considered as a fluid, subject to pressure forces and shocks.

Spiral structure is only transient in purely stellar disks.
In the absence of a cooling mechanism the disk is 
heated by the gravitational instabilities themselves, 
and reaches a hot steady-state, that prevents the development of
new instabilities. If a bar has developped, it
can remain for a Hubble time.
In presence of gas, even with as low a mass fraction as a few percents,
the disk is continuously cooled down, and responsive to new
dynamical instabilities. Due to its dissipative character, the gas acts as 
a regulator. If its conversion into stars and its continuous 
accretion is considered, the gas can prolong the phase of
 dynamical instability to more than a Hubble time, which can account
for the spiral density waves observed in so many spiral galaxies.

Dynamical instabilities play themselves a self-regulating role, since the bar
gravity torques drive the gas towards the center, and a sufficiently
massive central condensation begins to dissolve the bar.
Also, above a certain central gas concentration, a
one armed spiral structure can develop, which stops or slows
down the radial gas flow and the angular momentum transfer.
We have studied the conditions required for the m=1 mode appearance,
its pattern speed and extension. The m=1 perturbation is confined in the
central part of the galaxy, and is noticeable only within its 
own outer Lindblad resonance. Its angular speed is more than 10 times the 
speed of the external perturbations. It appears to have a purely gaseous
origin, and corresponds to the phenomenon identified in protostellar disks
by Adams et al. (1989). The mode grows in a few dynamical time-scales as 
predicted for the SLING mechanism of Shu et al. (1990). There is evidence 
from our simulations that stars and gas are off-centered with respect to the
center of mass of the system, which indicates that this mechanism
could be at work.

\begin{acknowledgements}
We are grateful to M. Tagger for useful discussions about the 
mode $m=1$. S.J. acknowledges financial support from the Brazilian 
CNPq (no. 201767-93.7), and wishes to thank the DEMIRM Department of
Observatoire de Paris for its hospitality.
\end{acknowledgements}

\end{document}